\documentclass[%
 reprint,
 superscriptaddress,
 amsmath,amssymb,
 aps,
prb,
floatfix,
longbibliography 
]{revtex4-2}

\usepackage[pdftex]{graphicx} \graphicspath{{}}
\usepackage{float}
\usepackage{ulem}
\usepackage{dcolumn}
\usepackage{bm}
\usepackage[pdftex,colorlinks=true]{hyperref}
\hypersetup{
  colorlinks=true,
  linkcolor=blue,
  urlcolor=cyan,
}
\usepackage[section]{placeins}

\newcommand{\eqa}[1]{\begin{align} #1 \end{align}}
\def\be{\begin{equation}}
\def\ee{\end{equation}}

\usepackage{color}

\newcommand{\amcr}[1]{\textsf{\color{red}(#1)}}

\newcommand{\nn}{\nonumber}

\newcommand{\mH}{\mathcal{H}}
\newcommand{\mD}{\mathcal{D}}
\newcommand{\mZ}{\mathcal{Z}}

\newcommand{\avg}[1]{\left\langle #1 \right\rangle}

\newcommand{\bS}{\boldsymbol{S}}

\newcommand{\ep}{\mathcal{E}}

\newcommand{\sectionn}[1]{\textit{#1---}}

\begin{document}

\preprint{APS/123-QED}

\title{Integrability-breaking-induced Mpemba effect in spin chains}%

\author{Adam J. McRoberts}
\affiliation{International Centre for Theoretical Physics, Strada Costiera 11, 34151, Trieste, Italy}
 
\date{\today}

\begin{abstract}
\noindent
We show that there are two distinct mechanisms that can cause the symmetry-restoration Mpemba effect in spin chains with \textit{weakly broken} integrability, such that the asymptotic equilibration is diffusive, but the lifetime of anomalously fast spin hydrodynamics at low temperature is parametrically large. In particular, we consider isotropic spin chains quenched out of equilibrium by suppressing the $z$-components, without inducing any net magnetisation. Initially, the restoration of isotropy is faster in hotter systems -- because they have more phase space available to scramble their initial conditions -- which may cause the equilibration curves to cross at early times in both integrable and non-integrable systems. At later times, however, the equilibration is effectively hydrodynamic, and the \textit{colder} systems start to equilibrate faster as the lifetime over which they evince superdiffusive spin hydrodynamics is parametrically larger -- but only in \textit{non}-integrable models. Depending on the details of the temperatures and the extent of the initial symmetry-breaking, two isotropy-restoration curves may have a crossing at early time, late time, neither, or both.
%
\end{abstract}

\maketitle

\sectionn{Introduction}
The Mpemba effect is one of the most famous apparent paradoxes in out-of-equilibrium physics, where a system initially further from equilibrium reaches the final state faster than a system that started closer. 
Brought into the modern era by the `misuse of a refrigerator'~\cite{mpemba1969cool}, the effect has been formulated more generally in dissipative stochastic systems~\cite{lu2017nonequilibrium,klich2019mpemba,bechhoefer2021fresh,teza2026speedups}, and been observed in a variety of physical contexts ranging from spin glasses~\cite{baity2019mpemba} to trapped ions~\cite{zhang2025observation} to colloidal systems~\cite{kumar2020exponentially}.

More recently, a different type of Mpemba effect has been described in the equilibration dynamics of closed, i.e., \textit{non}-dissipative systems. The Laws of the Game here are different, of course, because under Hamiltonian dynamics two ensembles will not generically equilibrate to the same final state; instead, one prepares the initial ensembles such that they break some symmetry of the Hamiltonian -- e.g., rotations generated by $\hat{S}^z$~\cite{rylands2024dynamical,rylands2024microscopic,murciano2024entanglement,chalas2024multiple,ares2025quantumfree,ares2025simpler} or lattice translation symmetry~\cite{klobas2025translation,gibbins2025translation} -- and measures the restoration of this symmetry under the dynamics.

This symmetry-restoration Mpemba effect has, up to now, most commonly been studied in integrable quantum models~\cite{ares2025quantum}; though it does not \textit{depend} on integrability, and has been observed in non-integrable systems~\cite{ares2023entanglement,joshi2024observing,liu2024symmetry,turkeshi2025quantum,yu2025tuning,yang2026probing}.
It is also known as the quantum Mpemba effect, though we will avoid using that term as the general name because it is not intrinsically quantum-mechanical, and occurs in classical systems as well.


The main result of this letter is that integrability-breaking \textit{itself} is a distinct mechanism for generating the symmetry-restoration Mpemba effect.

We first define the models, quench protocol, subsequent dynamical evolution, and the observable we use to measure the anisotropy; we then derive the analogue of the quantum Mpemba effect seen in, e.g., Refs.~\cite{rylands2024dynamical,murciano2024entanglement}. This manifests as early-time crossings of the equilibration curves in both integrable and non-integrable chains, and occurs because hotter systems generically restore the symmetry faster as they have more phase space to scramble their initial conditions. 

We then show that, for \textit{non-integrable chains only}, the parametrically longer lifetime of slower quasiparticles (solitons) reverses this generic expectation, comparatively speeding up the rate of equilibration in colder systems and leading to late-time crossings of the equilibration curves.

These two mechanisms, somewhat in opposition to each other, operate essentially independently; thus, depending on non-universal details, either, neither, or both crossings may occur.

\begin{figure*}[t]
    \centering
    \includegraphics[width=\linewidth]{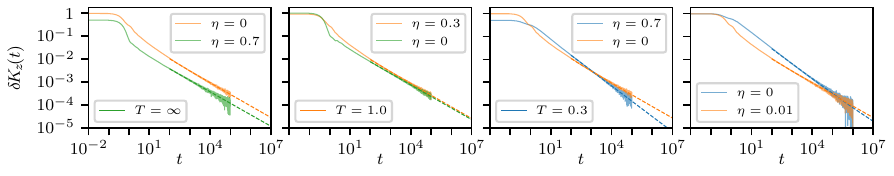}
    \caption{An example of each of the four possible scenarios for the equilibration curves, all for the classical Heisenberg chain~\eqref{eq:H_Heisenberg}. 
    Dashed lines are fits to the hydrodynamic form~\eqref{eq:hydro_eq} for $10^3 < t < 10^4$. From left to right: (I) No crossings -- the hotter chain starts closer to isotropy so there is no early-time crossing, and the colder chain crosses over to diffusive equilibration before the curves cross. (II) Early-time crossing only -- the hotter system starts further from isotropy but initially equilibrates faster; but there is no late time crossing because both chains cross over to diffusive equilibration relatively quickly. (III) Early- and late-time crossings -- the hotter chain starts further from isotropy and quickly crosses the colder equilibration curve; but the colder system equilibrates superdiffusively over a long enough timescale that there is a second crossing at late time. (IV) Late-time crossing only -- this is a clean demonstration of the integrability-breaking-induced Mpemba effect. The colder system starts further from isotropy and the early-time symmetry-restoration Mpemba effect does not occur; but the difference in the timescale associated to the superdiffusive equilibration is large enough that the curves cross at late time. In (I), (II), and (III) we use a system size of $L = 8192$ and ensembles of $10^5$ states; in (IV), we use $L = 32\,768$, and ensembles of $30\,000$ states.}
    \label{fig:Scenarios}
\end{figure*}

\vspace{0.2cm}
\sectionn{Models, initial ensembles, and quench protocol}
We consider in this letter the equilibration dynamics of isotropic classical spin chains. In particular, the integrable Ishimori chain~\cite{ishimori1982integrable},
\eqa{
\mH = -2J \sum_i \log \left(\frac{1 + \bS_i \cdot \bS_{i+1}}{2} \right),
\label{eq:H_Ishimori}
}
and the non-integrable classical Heisenberg chain,
\eqa{
\mH = -J \sum_i \left(\bS_i \cdot \bS_{i+1} -1 \right).
\label{eq:H_Heisenberg}
}
We can interpolate between these two models with the following spin chain,
\eqa{
\mH = - 2J \gamma^{-1} \sum_i \log \left(1 + \gamma \frac{\bS_i \cdot \bS_{i+1} - 1}{2} \right),
\label{eq:H_delta}
}
which is the Ishimori chain when $\gamma = 1$, and the Heisenberg chain when $\gamma \to 0$. We may regard $\delta = 1 - \gamma$, then, as the degree of integrability-breaking.

Note that we consider only \textit{ferromagnetic}, not antiferromagnetic chains, as only the former have long-lived spin superdiffusion~\cite{mcroberts2022anomalous,mcroberts2024parametrically,mcroberts2024anomalous}. From this point we set $J = 1$, with all temperatures measured in units of $J$ and times in units of $J^{-1}$, and use periodic boundary conditions throughout.

To probe the equilibration dynamics of these chains, we first construct an ensemble of thermal states at some temperature $T$ using heatbath Monte Carlo~\cite{loison2004canonical}, then force these states out of equilibrium by suppressing
the $z$-components by some factor $\eta$. In the canonical variables $\{\phi_i, z_i\}$, this is simply
\eqa{
S_i^x &= \sqrt{1 - z_i^2}\cos\phi_i \mapsto \sqrt{1 - \eta^2 z_i^2}\cos\phi_i \nn \\
S_i^y &= \sqrt{1 - z_i^2}\sin\phi_i \,\mapsto \sqrt{1 - \eta^2 z_i^2}\sin\phi_i \nn \\[0.1cm]
S_i^z &= z_i \mapsto \eta z_i.
}
Note that this does not induce any net magnetisation (though it will alter any residual finite-size magnetisation), and the in-plane components are appropriately re-scaled such that each spin remains normalised.

Subsequently, we let each state evolve under the dynamics of the Hamiltonian that generated the initial ensemble before the $z$-components were suppressed; the fundamental Poisson brackets are $\{\phi_i,z_j\} = \delta_{ij}$, and the equations of motion are
\eqa{
\dot{\bS}_i = 2\bS_i \times \left(\frac{\bS_{i-1}}{2 - \gamma + \gamma\bS_i\cdot\bS_{i-1}} + \frac{\bS_{i+1}}{2 - \gamma + \gamma\bS_i\cdot\bS_{i+1}}\right).
}

The suppression breaks the isotropy by changing the total weight of each of the spin components, which we denote by
\eqa{
K_{\mu}(t) = 3\avg{S_i^{\mu}(t)^2},
}
where the average is taken over both the sites $i$ and the states of the ensemble (though the ensemble average would not be necessary in the thermodynamic limit). The factor of $3$ is chosen such that isotropy means $K_\mu = 1$ for all $\mu$. We will use
\eqa{
\delta K_z(t) = 1 - K_z(t) 
}
to measure the degree to which isotropy is broken (that $\delta K_z = 0$ is both a necessary and sufficient condition to establish isotropy is discussed in the End Matter). Now, since the $K_\mu$ are not conserved quantities, the dynamical evolution will restore the isotropy, and we will compare the equilibration curves $\delta K_z(t)$ for different ensembles. 


We will permit the temperatures of the initial ensembles and the extent of symmetry-breaking $\eta$ to vary, but in all cases we will only compare ensembles generated by and evolving with the same Hamiltonian (i.e., a fixed value of $\gamma$ in Eq.~\eqref{eq:H_delta}). 

We say that the Mpemba effect occurs if, for two ensembles ($T_A$, $\eta_A$), $(T_B, \eta_B)$ with $\eta_A < \eta_B$ such that $A$ has a higher degree of initial symmetry-breaking, there exists some time $t^*$ such that for all $t > t^*$, $\delta K^A_z(t) < \delta K^B_z(t)$.

In the rest of this letter, we show that there are two mechanisms that can cause the equilibration curves $\delta K_z(t)$ to cross, and, because they act independently, they may cancel each other out.

\vspace{0.2cm}
\sectionn{Early-time crossings}
The first mechanism that can cause a crossing of the equilibration curves is that hotter systems have more phase space to scramble their initial conditions, and so, if the hotter ensemble is the one with a higher degree of symmetry-breaking, it may quickly overtake the equilibration curve of the colder ensemble. These early-time crossings can occur for both integrable and non-integrable systems.

If we consider only small $t$, we can straightforwardly calculate an initial equilibration rate by considering the Maclaurin series,
\eqa{
z_i(t) = z_i(0) + t\dot{z_i}(0) + \frac{t^2}{2}\ddot{z_i}(0) + ...\,.
}
This gives us
\eqa{
\delta K_z(t) = 1 - \eta^2 - 3\left(\avg{\dot{z}_i^2} + \avg{z_i\ddot{z}_i}\right)t^2 + ...,
}
where we omit temporal arguments when they are zero, and the linear term $\sim \avg{z_i\dot{z}_i}$ vanishes because the dynamics is time-reversible. We can simply take the coefficient of the quadratic term,
\eqa{
\Gamma = 3\left(\avg{\dot{z}_i^2} + \avg{z_i\ddot{z}_i}\right),
\label{eq:init_eq_rate}
}
as the early-time equilibration coefficient.

\begin{figure}[t]
    \centering
    \includegraphics[width=\linewidth]{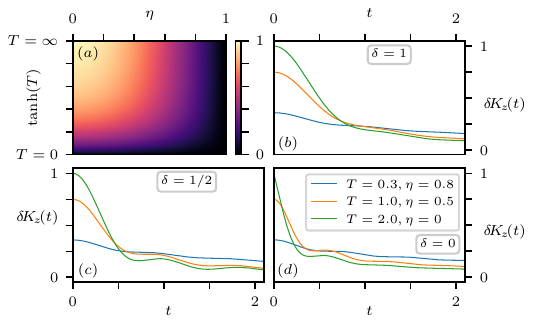}
    \caption{Early-time equilibration coefficients and crossings of the equilibration curves. (a) The initial equilibration rate $\Gamma(T, \eta)$ in the classical Heisenberg chain (cf. Eq.~\eqref{eq:init_eq_rate}), showing that $\Gamma$ is not purely a function of $\eta$, and thus a system with an initially higher degree of symmetry-breaking may equilibrate faster than a system with lower initial symmetry-breaking. 
    (b), (c), and (d) show example crossings of the equilibration curves for the Heisenberg chain, $\gamma = 0.5$ chain (cf. Eq.~\eqref{eq:H_delta}), and Ishimori chain, respectively. In some cases, the oscillations can even cause multiple crossings, as seen in (d). }
    \label{fig:Short-time}
\end{figure}

We will write out explicit calculations only for the Heisenberg chain; the extension to the interpolating model \eqref{eq:H_delta} is straightforward, though somewhat tedious. 
First, we have, from the equations of motion,
\eqa{
\dot{z}_i = \sqrt{1 - z_i^2}\,\Bigl(&\sqrt{1 - z_{i+1}^2}\sin(\phi_{i+1} - \phi_i) \nn \\
&- \sqrt{1 - z_{i-1}^2}\sin(\phi_i - \phi_{i-1})\Bigr).
}
Without loss of generality, we may fix the site index to $i = 0$. Recalling that what we want is to calculate $\avg{\dot{z}_0^2}$ \textit{after} $z \mapsto \eta z$, we thus have
%
%
\eqa{
\avg{\dot{z}_0^2} = \frac{1}{\mZ^{(2)}}&\int d\Omega_{-1} d\Omega_0\,d\Omega_1\,e^{\beta J(\ep_{-1,0} + \ep_{0,1})} \left(1 \!-\! \eta^2 z_0^2\right) \nn \\
&\times \biggl(\sqrt{1 - \eta^2 z_{1}^2}\sin(\phi_{1} \!-\! \phi_0) \nn \\[-0.2cm]
&\;\;\;\;\;\;\;\;- \sqrt{1 - \eta^2 z_{-1}^2}\sin(\phi_0 \!-\! \phi_{-1})\biggr)^2,
\label{eq:init_eq_rate_H_term_1}
}
where $d\Omega := dz\,d\phi/(4\pi)$ is the integration measure; $\ep_{i,i+1}$ is the energy of a bond \textit{before} the isotropy was broken, 
\eqa{
\ep_{i,i+1} = z_i z_{i+1} + \sqrt{1 \!-\! z_i^2}\sqrt{1 \!-\! z_{i+1}^2}\cos(\phi_i \!-\! \phi_{i+1});
 } 
and $\mZ^{(2)}$ is the partition function restricted to the two relevant bonds,
\eqa{
\mZ^{(2)} = \int d\Omega_{-1} d\Omega_0\,d\Omega_1\,e^{\beta J(\ep_{-1,0} + \ep_{0,1})}.
}
The integral in Eq.~\eqref{eq:init_eq_rate_H_term_1} cannot be done analytically (except at $T = \infty$), though it is straightforward to evaluate it numerically using heatbath Monte Carlo: simply construct a large ensemble of exact thermal states on the three sites $-1$, $0$, and $1$; map $z \mapsto \eta z$ and re-scale the in-plane components; then explicitly calculate $\dot{z}_0^2$ for each out-of-equilibrium state and take the average.

The second term in Eq.~\eqref{eq:init_eq_rate} involves $\ddot{z}$, the full expression for which involves spin components on five sites. Again, the integrals cannot be done analytically away from $T = \infty$, but the numerical procedure is the same (except, of course, that the thermal states should be constructed on five sites).

\begin{figure}[!t]
    \centering
    \includegraphics[width=\linewidth]{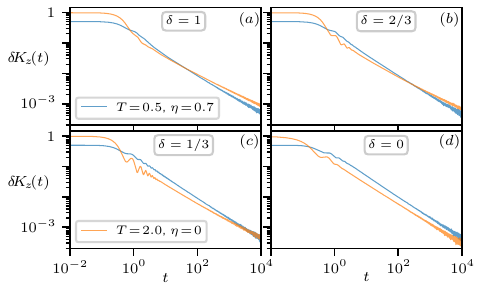}
    \caption{Dependence of the crossing times on the degree of integrability-breaking $\delta$, in the case where there are both early- and late-time crossings. The early-time crossing depends only very weakly on $\delta$, though there are pronounced oscillations of $\delta K_z(t)$ as the integrable point $\delta = 0$ is approached. The late-time crossing depends very strongly on $\delta$, occurring (for these temperatures and quench parameters) around (a) $t^* \approx 10^2$ for $\delta = 1$ (the Heisenberg chain); (b)  $t^*\approx 10^3$ for $\delta = 2/3$; and (c) $t^* \approx 10^4$ for $\delta = 1/3$. There is no late-time crossing for (d) $\delta = 0$ (the Ishimori chain), as integrability means the hydrodynamic equilibration is perfectly superdiffusive at all temperatures even as $t \to \infty$. In all cases here we use a system size of $L = 8192$ and average over $10^5$ states.}
    \label{fig:double_crossings}
\end{figure}

Fig.~\ref{fig:Short-time}(a) shows the early-time equilibration coefficients $\Gamma(T, \eta)$ for the Heisenberg chain, numerically calculated using ensembles of $10^6$ states exactly sampled from the Gibbs distribution for five-site chains. Clearly, it is generically possible to select values of $(T_A, \eta_A)$ and $(T_B, \eta_B)$ such that ${\eta_A < \eta_B}$ (i.e., $A$ initially has a higher degree of symmetry-breaking than $B$), but ${\Gamma_A > \Gamma_B}$, so $A$ equilibrates at a faster rate than $B$ and the symmetry-restoration curves can cross. 

Several examples of these short-time crossings are shown in Figs.~\ref{fig:Short-time}(b), (c), \& (d), for the Heisenberg chain, the $\gamma = 0.5$ chain (cf. Eq.~\eqref{eq:H_delta}), and the Ishimori chain, respectively. Depending on the details, the oscillations in the equilibration curves can lead to multiple crossings, as seen in Fig.~\ref{fig:Short-time}(d).

These crossings of the symmetry-restorations curves are the direct classical counterpart of the quantum Mpemba effect observed in systems such as Refs.~\cite{rylands2024dynamical,murciano2024entanglement}.

\vspace{0.2cm}
\sectionn{Integrability-breaking-induced Mpemba effect}
The second mechanism that can cause the equilibration curves $\delta K_z(t)$ to cross is a separation of hydrodynamic timescales as a function of temperature; this can occur only in \textit{non-integrable} chains, and thus we dub this the integrability-breaking-induced Mpemba effect.

\begin{figure}[!t]
    \centering
    \includegraphics[width=\linewidth]{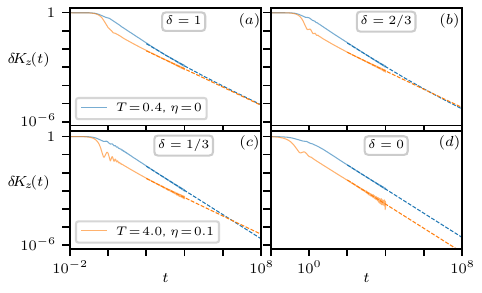}
    \caption{The integrability-breaking-induced Mpemba effect. (a) For these temperatures and quench parameters the crossing has disappeared at $\delta = 1$ (the Heisenberg chain), but is present for (b) $\delta = 2/3$ and (c) $\delta = 1/3$, and the Mpemba effect is observed in these cases. (d) The crossing is forbidden in the integrable case $\delta = 0$ (the Ishimori chain). The curves are extrapolated by fitting the hydrodynamic form Eq.~\eqref{eq:hydro_eq} between $t = 3000$ and $t = 10\,000$; the validity of this extrapolation can be inferred from the agreement between the curves for $100 < t < 3000$, where they are not fitted, implying we have indeed entered the hydrodynamic equilibration regime after $t \approx 100$. In all cases here we use a system size of $L = 8192$ and average over $10^5$ states.}
    \label{fig:late_crossings}
\end{figure}

Although the weights of the various components $K_\mu$ are not conserved quantities, establishing the new global equilibrium after the quench requires the transport of conserved quantities over long distances, and thus the late time relaxation to isotropy is hydrodynamic~\cite{lux2014hydrodynamic}. In the integrable Ishimori chain with KPZ-like spin superdiffusion~\cite{das2019kardar}, this means $\delta K_z(t) \sim t^{-2/3}$ at late times.

Asymptotically, non-integrable spin chains have diffusive spin correlations, and so the late time equilibration in those chains follows a diffusive power-law,
\eqa{
\delta K_z(t) \propto 1/\sqrt{D t}, \;\;\; t\to\infty,
}
where $D$ is the diffusion constant. However, whilst this \textit{is} asymptotically correct, the solitons that are responsible for superdiffusion at the integrable point are adiabatically connected to solitons in the non-integrable chain~\cite{mcroberts2022long}, and, particularly at low temperature or near the integrable point (small $\delta$), source a parametrically long-lived regime of superdiffusion~\cite{mcroberts2022anomalous,mcroberts2024parametrically,mccarthy2024slow,wang2025breakdown}.

The effective diffusion constant as a function of time in isotropic spin chains, therefore, has the form~\cite{mcroberts2024parametrically}
\eqa{
D(t) \sim \frac{\mD}{1 + \lambda t^{-1/3}},
}
where the parameter $\lambda$ controls the timescale $\tau \sim \lambda^3$ for the crossover from superdiffusive to diffusive hydrodynamics. This timescale diverges like $\tau \sim \delta^{-3}$ as a function of integrability-breaking, and $\tau \sim T^{-8}$ as a function of temperature~\cite{mcroberts2024parametrically}. Incorporating this long-lived superdiffusive regime, an improved form for the hydrodynamic equilibration curves is
\eqa{
\delta K_z(t) \sim \kappa \sqrt{\lambda t^{-4/3} + t^{-1}},
\label{eq:hydro_eq}
}
for some constant $\kappa$. Consider, then, two ensembles $A$ and $B$ in the hydrodynamic regime; the late-time crossing, if there is one, occurs at
\eqa{
t^* = \left(\frac{\kappa_A^2\lambda_A - \kappa_B^2\lambda_B}{\kappa_B^2 - \kappa_A^2}\right)^3.
\label{eq:late_crossing_time}
}
The condition that they actually do cross is simply that Eq.~\eqref{eq:late_crossing_time} returns $t^* > 0$, and, because the equilibration has reached the hydrodynamic regime, this is always the last crossing. We show examples of these late-time crossings, and their dependence on the degree of integrability-breaking, in Figs.~\ref{fig:double_crossings} \& \ref{fig:late_crossings}. 

The case where there is a crossing at early-time and at late-time, with the Mpemba effect essentially cancelled out, is shown in Fig.~\ref{fig:double_crossings}; as integrability is approached, however, the late-time crossing is pushed ever further back (Fig.~\ref{fig:double_crossings}(a)-(c)), until it is entirely prevented -- and the Mpemba effect from the early-time crossing restored -- in the integrable chain (Fig.~\ref{fig:double_crossings}(d)). 

The case where the late-time crossing occurs \textit{without} an early-time crossing, and therefore itself constitutes an Mpemba effect, is shown in Fig.~\ref{fig:late_crossings}, using equilibration curves extrapolated from the hydrodynamic form~\eqref{eq:hydro_eq}. The value of integrability-breaking $\delta = 1 - \gamma$ that results in the fastest crossing (when there is one) for specific temperatures and symmetry-breaking is not universal -- if $\delta$ is too large, both curves will reach the diffusive regime before they can cross; if $\delta$ is too small, the crossover from superdiffusive to diffusive equilibration will be too slow for the curves to cross. 

This integrability-breaking-induced Mpemba effect is reminiscent of, but is not quite, the strong Mpemba effect~\cite{klich2019mpemba}; for hydrodynamic relaxation that is usually defined in terms of the \textit{asymptotic} equilibration occurring with two different power-laws. In this case, the truly asymptotic equilibration of any non-integrable spin chain is \textit{always} diffusive -- but this fact alone ignores the parametrically long timescales over which effectively superdiffusive equilibration is observed. 

Nevertheless, it is the very fact that there \textit{is} eventually a crossover to diffusion that causes the integrability-breaking-induced Mpemba effect, by slowing the equilibration of hotter systems; the hotter systems have a higher density of fast, high-energy solitons, leading to initially faster equilibration -- but, when integrability is broken, these are the quasiparticles with the shortest lifetimes, and it is the colder systems that have a larger density of what are now parametrically faster modes.






\vspace{0.2cm}

\sectionn{Conclusions}
In this letter, we have shown that there are two independent mechanisms that can cause the equilibration curves of spin chains to cross and exhibit the Mpemba effect -- unless both crossings occur and, in some sense, cancel each other out.

The first mechanism is a classical analogue of the quantum Mpemba effect observed in integrable systems, where the scrambling of initial conditions is faster in hotter systems; and the second is a hydrodynamic effect that can occur only in \textit{non-integrable} spin chains, where the long-lived superdiffusive regime causes \textit{colder} systems to equilibrate with a faster effective power-law over long (but finite) timescales. 

Integrability is prized as the rarefied structure on which exact solutions can be built; but it can never be perfectly realised in practice, and the question of how approximate integrability is evinced is clearly of practical and experimental importance. One aspect of this work is to highlight that the breaking of integrability can itself be intrinsically interesting and lead to new phenomena. 

The Mpemba effect, broadly defined as a crossing of equilibration curves, is by now understood to occur in an array of physical systems by a variety of mechanisms, and is one manifestation of the richness of physics out of equilibrium. Whether integrability-breaking can lead to the observation of the Mpemba effect in other physical settings, via temperature- or energy-dependent quasiparticle lifetimes, is a particularly intriguing question.

\vspace{0.2cm}
I am grateful to C. A. Hooley and R. Moessner for helpful discussions, and in particular to C. Rylands and {M. Gibbins} for their comments on the draft. 



\bibliography{refs} 

\onecolumngrid
\vspace{1cm}

  \begin{center}
    \textbf{\large End Matter}
  \end{center}
\setcounter{equation}{0}
\setcounter{figure}{0}
\setcounter{table}{0}
\makeatletter
\renewcommand{\theequation}{S\arabic{equation}}
\renewcommand{\thefigure}{S\arabic{figure}}
\renewcommand{\thetable}{S\arabic{table}}
\setcounter{section}{0}
\renewcommand{\thesection}{S-\Roman{section}}

\setcounter{secnumdepth}{2}

\vspace{0.5cm}

\twocolumngrid


In this appendix, we explicitly justify the use of $\delta K_z(t)$ as a measure of the symmetry-breaking. In principle, of course, $\delta K_z = 0$ is only a necessary, but not sufficient condition to establish isotropy, and we ought to consider the full probability density function $\rho(S^z)$. In practice, however, given the way the distribution $\rho(S^z)$ actually evolves under the Hamiltonian dynamics, $\delta K_z = 0$ is also sufficient. 

If isotropy is unbroken, then each $\bS_i$ is uniformly distributed over the sphere, and the distribution of the $z$-components is simply the uniform distribution on the interval $[-1, 1]$; the initial out-of-equilibrium ensemble has $z$-components drawn from the uniform distribution on the interval $[-\eta, \eta]$.

Now, one could imagine that, at some finite time during the evolution, the variance $\avg{z^2}$ might attain its equilibrium value of $1/3$ (and thus $\delta K_z = 0$) but the other (even) moments, say, could be wrong, and thus the system would still be anisotropic. One ought, then, to define some proper distance between the numerically sampled distribution $\rho_t(z)$ at time $t$ and the equilibrium distribution $\rho_\infty(z) = 1/2$.

\begin{figure}[t]
    \centering
    \includegraphics[]{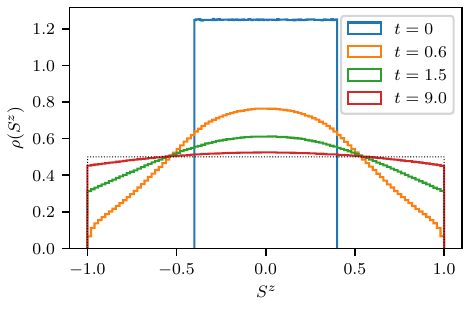}
    \caption{Evolution of the full one-spin probability distribution function $\rho_t(S^z)$ of the $z$-components, evolving from the initial out-of-equilibrium uniform distribution over the interval $[-\eta, \eta]$ towards, asymptotically, the equilibrium uniform distribution over the interval $[-1, 1]$. At all times, we observe, $\rho_t(z)$ is a non-increasing function of $|z|$; thus, at all times, the variance $\avg{z^2} < 1/3$ is less than the equilibrium value. Indeed, all the even moments $\avg{z^{2n}} < 1/(2n+1)$ are below their equilibrium values. It follows that $\avg{z^2}$ only attains its equilibrium value of $1/3$ if $\rho_t(z)$ attains the equilibrium distribution, which justifies the use of $\delta K_z(t)$ as a measure of anisotropy. In this figure, $T = 1.0$, $\eta = 0.4$, the system size is $L = 8192$, and we use an ensemble of $10^4$ states. The dotted line shows the equilibrium distribution.}
    \label{fig:histograms}
\end{figure}

But, if we examine the full distribution $\rho_t(z)$, cf. Fig.~\ref{fig:histograms}, we see that, for any time $t$, $\rho_t(z)$ is a non-increasing function of the absolute value $|z|$, and converges to $\rho_\infty$ asymptotically (though, clearly, this statement is only supported by numerical evidence, and not mathematically proven).

Thus, the variance $\avg{z^2} \to 1/3$ only if $\rho_t \to \rho_\infty$, and the possibility that one could attain $\delta K_z = 0$ without 
the 
\vfill\eject
\noindent proper restoration of isotropy does not occur; which justifies its use as the measure of symmetry-breaking.


\end{document}